\documentstyle[preprint,prb,aps,psfig]{revtex}
\begin{document} 
\draft
\narrowtext
\title{\Large\bf A Model for the
Thermal Expansion of Ag(111) and other Metal Surfaces }
\author{Shobhana Narasimhan and Matthias Scheffler}
\address{
Fritz-Haber-Institut der Max-Planck-Gesellschaft,\\ Faradayweg 4-6,
D-14195 Berlin-Dahlem, Germany}

\date{\today}
\maketitle
\begin{quote}
\parbox{16cm}{\small


We develop a model to study  the thermal expansion of surfaces, wherein  phonon
frequencies are
obtained from {\it ab initio} total energy calculations. Anharmonic effects
are treated exactly in the direction normal to the surface, and
within a quasiharmonic approximation in the plane of the surface. We apply this model to the Ag(111) and Al(111)
surfaces, and find that our calculations
reproduce the experimental
observation of a large and anomalous increase in the surface thermal expansion of Ag(111)
at high
temperatures\cite{statiris94}. Surprisingly, we find
that this increase can be attributed to a rapid softening
of the {\it in-plane}
phonon frequencies, rather than due to the anharmonicity of the out-of-plane surface phonon modes. This
provides evidence for a new mechanism for the enhancement of surface anharmonicity. A comparison with Al(111)
shows that the two surfaces behave quite differently, with no evidence for such anomalous 
behavior on Al(111).
}
\end{quote}

The equilibrium lattice constant of a crystal is determined by 
a balance between the various attractive and repulsive forces 
present within the solid. When the crystal is cleaved to form surfaces, this
balance is destroyed, and the atoms at the surface therefore relax
either
inwards (which is the usual situation for most metal surfaces) or outwards. 

However, upon heating the crystal, these relaxations may change dramatically
as a function of temperature. The phenomenon of bulk thermal expansion is, of
course, a familiar one: upon heating the crystal, the lattice constant changes,
reflecting the anharmonicity of the interatomic potentials. Similarly, asymmetries
and anharmonicities in surface phonon vibrations will be manifested in a change
in interlayer displacements as a function of temperature $T$. Due to the
different asymmetries present at the surface and in the bulk, it is possible
that these two quantities (bulk lattice constant and interlayer distances near
the surface) may change quite differently upon increasing $T$. 

Indeed, it has
long been realized that one would expect all such measures of anharmonicity 
(e.g., coefficients of thermal expansion $\alpha$,
mean squared displacements (MSDs) of atoms, and the rate of change of
phonon frequency with temperature) to be larger at surfaces than in bulk crystals\cite{allen69}
for two reasons: 
{\it (i)} the breaking of symmetry due to the presence of
the surface makes the interlayer potential more asymmetric at
the surface than in the bulk, increasing the size of the
odd terms in
a Taylor series expansion
of the energy in powers of atomic displacements; 
{\it (ii)} atomic displacements may be
greater at the surface than in the bulk, thus increasing the
relative magnitudes of the higher-order anharmonic terms in this series expansion. 

Interestingly, studies on metal surfaces have shown that such enhancements of surface anharmonicity,
when observed, are strongly dependent on both the element and the orientation of the surface.
Early experiments and calculations
\cite{wilson71},\cite{dob73ma78} on metal
surfaces suggested that  measures of anharmonicity (such as the coefficient of thermal expansion) 
are
typically at most three times larger at the surface than in the bulk, and as a result do not
affect surface properties drastically. However, recent experiments have
shown a huge enhancement on a  few surfaces:  for Ni(001)\cite{cao90}
and Pb(110)\cite{frenken87} at high
temperatures, the surface coefficient of thermal expansion $\alpha_S$, defined
by:

\begin{equation}\label{eq:alphas}
\alpha_S  = (d_{12})^{-1}(\partial d_{12}/\partial T) , 
\end{equation}

\noindent
(where $d_{12}$ is the interlayer spacing between the first two planes of atoms
at the surface)
is 10 to 20 times larger than the bulk coefficient of thermal expansion
$\alpha_B$, while for Cu(110) the surface MSDs are up to six times greater than
the bulk MSDs\cite{helgesen93}.

Perhaps the most interesting case is that of Ag(111): recent experiments \cite{statiris94} show that
the contraction of $d_{12}$ is dramatically reversed for the Ag(111) 
surface as the crystal is heated:
Up to 
$T \approx 670$ K,  $d_{12}$ is indeed
{\it contracted} by $\sim$ 2.5\% relative to the bulk separation
$d_B$; but upon increasing $T$ further, $d_{12}$ increases much more rapidly than
$d_B$ does, so that by 1150 K it is {\it expanded} by $\sim$ 10\%. Correspondingly,
$\alpha_S$ becomes more than ten times as large as
$\alpha_B$ \cite{explain}. 
Such a large effect is especially unexpected for a close-packed metal surface, and
contradicts the conventional
expectation that more open surfaces should exhibit larger surface anharmonicity.

Lewis\cite{lewis95} has carried out EAM simulations  to investigate the thermal behavior of Ag(111). The
results of these simulations differ significantly from those reported
experimentally: the surface layer relaxes inwards at all temperatures,
and $\alpha_S$ is less than twice as large as $\alpha_B$.

Is this disagreement between experiment and calculation on Ag(111) due to
inadequacies of the EAM potentials? Or could it be a sign of some hitherto
undetected surface phase transition?

To study these questions,
we have investigated the harmonic and anharmonic properties of Ag(111) and Al(111) 
by performing {\it ab initio} calculations using density functional theory.
Fully separable norm-conserving pseudopotentials\cite{gonze91}
were used in our calculations, together with a plane wave
basis set with an energy cut-off of 60 Ry (20 Ry for Al), and  the local-density
approximation with Ceperley-Alder exchange-correlation\cite{ceperley80}. We note that the
relatively high cut-off of the plane-wave basis set is necessary in order to obtain a
good description of {\it anharmonic} effects, even though a lower cut-off may suffice to
describe harmonic properties adequately.

Before performing surface calculations, we first verified that the harmonic and
anharmonic properties of the bulk materials are satisfactorily described by these
pseudopotentials. Some of these results, such as the optimal values of the lattice 
constant and the bulk modulus, are presented in Table 1. This table also contains
the calculated values of two measures of anharmonicity: the pressure derivative
of the bulk modulus, and the Gr\"uneisen parameters $\gamma$, which describe how the
phonon frequencies vary upon changing the lattice constant. Note that the experimental value
of $\gamma$ is an average over all bulk modes, whereas the theoretical values were
calculated separately for each band, and at a sample wave-vector along the [111] direction, between the
zone-center $\Gamma$ and the zone-edge $L$. (These particular bulk modes were chosen because
they project on to the zone-center of the surface Brillouin zone, and can 
therefore be regarded as analogous to the surface zone-center vibrations that
we will later investigate for the (111) surface.)

The surface calculations were performed using a repeated slab geometry consisting of six
atomic layers separated by a vacuum layer of the same thickness. 
The {\bf k}-point sets used to sample reciprocal space consisted of a
uniform grid centered on the $\Gamma$ point and containing seven points
in the irreducible part of the Brillouin zone for the undistorted  surface;
the number of {\bf k}-points was correspondingly increased upon breaking symmetries 
by distorting the lattice in order to calculate phonon frequencies.  Convergence
of calculated anharmonic quantities
with respect to energy cut-off, number of {\bf k}-points and number of layers was carefully
tested for.

Our strategy is to compute static energies and phonon frequencies by performing
self-consistent calculations at $T=0$ K, and then extend our results to finite temperatures
by using a quasiharmonic approximation.

In order to obtain a qualitative understanding of the mechanisms in operation, and a first estimate
of the size of surface thermal effects,
we use a simple model
of the lattice dynamics of the surface, considering only three phonon modes, in all of
which the topmost layer of the slab moves as a whole -- i.e., we assume that the
displacements are confined to the first layer of atoms at the surface, and consider
only those modes with zero wave-vector. We consider one mode in which the surface
atoms vibrate normal to the surface plane (along the $z$ direction), and two modes
(along ${\bf x}$ = $[1 { \overline1} 0]$ and ${\bf y}$ = $[1 1 {\overline2}]$)
in which they vibrate in the
plane of the surface. ( We emphasize that these displacements 
do
not correspond to any of the actual normal modes of vibration of the surface slab,
but may be considered as indicative of the strengths of the various force-constants
that would appear in the true dynamical matrix of the system. This proviso should
be kept in mind when we refer to ``modes" and ``phonons" in the rest of this
paper.)

We first computed the change in the total
energy of the Ag(111) slab upon varying the first two interlayer separations $d_{12}$ and
$d_{23}$, and maintaining the fcc stacking of the bulk crystal.
This not only provides the static interlayer potential, but is also equivalent 
to simulating the vibrational mode along $z$. Our result for the dependence on $d_{12}$ of the 
first interlayer potential is plotted in Fig.~1; it is clearly asymmetric about
the minimum at $d_{12} = 2.30$ \AA. We found that
allowing for the relaxation of  $d_{23}$ does not have a significant impact on the
results for the close-packed (111) surface\cite{d23}.
 For
the results presented in Fig.~1 and the rest of this paper, $d_{23}$ is therefore fixed at the bulk
interlayer separation of $2.34$ \AA.

To see how this anharmonicity of the interlayer potential is manifested at finite
temperatures, we consider a one-dimensional quantum oscillator vibrating in the interlayer
potential of Fig.~1. A numerical solution of the Schr\"odinger equation for this problem furnishes
the eigenstates and eigenvalues of such an oscillator, and  the mean displacement
$\langle d_{12}\rangle_n$  in the $n$-th eigenstate is obtained by computing the
expectation value of the displacement operator in each state. The average value at a
finite temperature $T$ is then obtained by weighting these results with the corresponding
partition function.

Our results for
$d_{12}(T)$ obtained from this procedure (see the open
circles in Fig.~3) indicate a 
modest enhancement in $\alpha_S$ relative to $\alpha_B$ of $\sim 1.7$, which is  much
smaller than that measured experimentally. 


%
%

We next performed frozen-phonon calculations
to study the behavior of the two in-plane modes
in our model. At each value of  $d_{12}$, the atoms in the surface layer were displaced along
first the $x$ and then the $y$ direction, and the total energy was computed for a series of
displacements up to $\pm 0.15$ \AA. The curvature of the resulting plots of energy versus 
displacement gives the mode frequency. 
We find that the frequency of these in-plane modes 
decreases surprisingly rapidly upon increasing $d_{12}$; these results are
plotted in Fig.~2. 

In analogy to the usual bulk Gr\"uneisen parameter $\gamma_B$, we can define a surface 
Gr\"uneisen parameter:

\begin{equation}\label{eq:gammas}
\gamma_S \equiv -\partial {\rm ln}(\omega)/\partial {\rm ln}(d_{12}).
\end{equation}
\noindent
The value of the $\gamma_S$ extracted from our results for the in-plane modes (between 
5.5 and 7.5) is significantly larger than  the corresponding value of of 2.39 that
we obtain for a bulk vibration with an analogous pattern of displacements.
Thus, the frequently made assumption\cite{wilson71} that $\gamma$ is approximately equal for surface and bulk modes
is clearly invalid in this case.

The surface can thus reduce its vibrational free energy significantly by expanding outwards, though
such
an outward expansion would be accompanied by an increase in the static energy. The optimal
value of $d_{12}$ is determined by minimizing the free energy
function\cite{allen69}:

\begin{equation}\label{eq:ftot}
F(d_{12},T)  = E_{\rm stat}(d_{12}) + \sum_{i} F_{\rm vib}^{i}(d_{12},T) , 
\end{equation}
\noindent
at each temperature $T$ . Here, $E_{\rm stat}(d_{12})$ is the static interlayer potential plotted in
Fig.~1, and $F_{\rm vib}^i(d_{12},T)$ is the vibrational free energy corresponding to vibrations in
the $i$-th direction, which, in the quasiharmonic approximation, is given by\cite{allen69}:
\begin{equation}\label{eq:fvibqh}
F_{\rm vib}^{i}(d_{12},T) = k_B T ln \Bigl\{2 sinh \Bigl({{\hbar\omega_i(d_{12})}\over{2 k_B T}}\Bigr)\Bigr\};
\end{equation}
\noindent
where $k_B$ and $\hbar$ are Boltzmann's constant and Planck's constant respectively.
The frequency of the mode polarized along the $i$-th direction, when the first interlayer
separation is fixed at $d_{12}$, is denoted by $\omega_i(d_{12})$. The sum in
Eq. (3) runs over all the bands of the phonon spectrum (averaged over the 
entire surface Brillouin zone); in our case we approximate
it by a sum over the 3 zone-center vibrational patterns that we have considered,
which are polarized along the $x$,$y$ and $z$ directions respectively. 
The variation of $\omega_i(d_{12})$  is obtained directly from our frozen-phonon calculations for the
two in-plane modes (see Fig.~2). For the out-of-plane mode, we compute $F_{\rm vib}^z(d_{12},T)$ numerically, in such a
way as to reproduce the exact result for $d_{12}(T)$ that we have already obtained by solving the
Schr\"odinger equation when only the mode along $z$ is present.

Our final result for $d_{12}(T)$ in the presence of all three modes, obtained by minimizing the
free energy expression given by Eq.~(3), is shown by the filled circles in Fig.~3 (note that the temperatures
are normalized with respect to the melting temperature $T_m$).
The experimentally measured data points\cite{statiris94} are also plotted; both the 
experimental and theoretical curves display the same features:
there is little or no change up to about $T/T_m = 0.5$, i.e., $T = 617$ K, 
beyond which there is a rapidly increasing trend towards
outwards relaxation of the surface layer. 
At $T/T_m = 0.85$, i.e., $T = 1049$ K, we find that the surface layer is relaxed outwards
by about 15\%, whereas 
the experiments show an outwards relaxation of $\sim 7.5\%$. Given the simplicity of our model,
and the large experimental error bars, this is as good an agreement as we can hope for.

The increasing slope of $d_{12}(T)$ reflects a flattening
in the minimum of the free-energy curve, and the rapid increase in $d_{12}(T)$
at high $T$ is a precursor to the development of a saddle-point 
instability in the free-energy curve, similar to that which has
been obtained in studies of the surface melting of copper surfaces
\cite{jayanthi}.  We emphasize that the rapid decrease of $\omega_i(d_{12})$ for the in-plane
modes is crucial for obtaining the large outwards expansion; if, for example, this rate of decrease
were to be halved,
the maximum outwards expansion would be  drastically reduced to
about 2\%. In the limit of high $T$ (when all modes are excited), the value of $d_{12}(T)$ is
no longer sensitive to the absolute scale of $\omega_i$, but is instead controlled by $\gamma_S$,
which is a normalized indicator of how rapidly $\omega_i$ falls off with increasing $d_{12}$.

To check whether such behavior is universal or a peculiar property of Ag(111),
we repeated the same calculations
on bulk aluminum and Al(111). Our results for the relaxation of $d_{12}$ for Al(111) are also plotted in
Fig.~3, and it is obvious that there is no evidence for a dramatically increased surface expansion
on Al(111). We have also performed calculations on Cu(111)\cite{unpub} which show that the behavior
of Cu(111) is intermediate between that of Ag(111) and Al(111), which has been confirmed by very
recent experiments \cite{gustpriv}.

Our results indicate that in addition to the two well known  sources of enhanced surface
anharmonicity that we have already mentioned, a third
(and hitherto neglected)
effect is more important in causing
the dramatic enhancement in surface
anharmonicity on Ag(111): Not only do interlayer potentials at the surface tail off
rapidly with increasing $z$ , they
simultaneously {\it become much flatter in the xy plane} -- in other
words, the operative effect is not so much a decrease in the absolute magnitudes
of interlayer
potentials at the surface, but a {\it reduction in their corrugation parallel to
surface}. As a consequence, those surface phonon modes in which atomic
dispacements have significant components in the surface plane soften rapidly
upon increasing interlayer separations; this drives the outermost layer of atoms to
expand outwards at high temperatures.

An
accurate description of the in-plane corrugation clearly requires taking into account the 
correct distribution of the electronic charge density (and the resulting chemical bond
formation) at the surface. It seems plausible that when rebonding effects become
significant, the EAM (which essentially ignores the relaxation of atomic charge densities
and the rehybridization of electronic states) may fail; this may be why Lewis \cite{lewis95}
did not observe
a large outwards expansion in his  simulations.

The rapid decrease in the corrugation of the interlayer potential is evident in Fig.~4, 
where we have plotted the differences
in energies when the outermost layer of atoms occupies various stacking sites. Note that
{\it (i)} the flattening occurs more rapidly for Ag(111) than Al(111) {\it (ii)} upon allowing for
the lighter mass of Al atoms, the {\it effective} corrugation relevant for phonon
frequencies is actually larger for Al(111) than for Ag(111).  Both these factors contribute
to the enhancement in $\gamma$ of the in-plane top-layer modes and thus the larger thermal expansion of Ag(111).

Such a decrease in the corrugation of the substrate potential would also tend to favor a top-layer
reconstruction of the type that has been observed on Au(111)\cite{au111expt} or Pt(111)\cite{pt111expt},
where the substrate potential is too weak to prevent a densification of atoms in the topmost layer.
However,  experiments
apparently show no evidence of such a reconstruction on Ag(111)\cite{statiris94}; further calculations of surface stresses and the
strength of intralayer couplings should help clarify the situation.

Our conclusion that the enhancement in surface expansion arises mainly from in-plane vibrations is supported
by the experimental observation that the amplitude of in-plane vibrations on Ag(111) rises faster than the 
magnitude of out-of-plane vibrations\cite{gustpriv}. 
Experiments on other surfaces, e.g., Cu(001), have detected in-plane vibrational
amplitudes that are larger than out-of-plane amplitudes\cite{jiang91},
and it is interesting to speculate whether this
(counter-intuitive) result arises from the same cause, i.e., from a rapid softening of in-plane frequencies due to
thermal expansion. 

Of course, the calculation we have presented above does contain certain
approximations: we have restricted ourselves to surface vibrations at the
zone-center, we approximate the true electronic exchange-correlation functional
by the LDA, and we use a quasiharmonic form for vibrational free energy.
We should also allow 
for expansion in the plane of the surface, but this should in fact reinforce the 
softening of the in-plane modes. 
The validity of some of these approximations can be tested by performing {\it ab initio} molecular dynamics
simulations in which all anharmonic contributions are fully included automatically, and in which no
assumptions are made about the polarizations of bulk and surface vibrations; work in this direction
is in progress.

We expect that our numerical results will change slightly upon including other surface phonon modes and
allowing for dispersion through the surface Brillouin zone.  Surface vibrations at the zone-center are
sensitive only to the strength of {\it interlayer} force-constants coupling atoms at the surface
to atoms in layers below; however, the frequency of a phonon at arbitrary
wave-vector also depends on the {\it intralayer} force-constants, and one may
expect these to be less sensitive to $d_{12}$; we may therefore have over-estimated
the degree of surface anharmonicity by restricting ourselves to the zone-center. 
However, we note that measurements of surface phonon frequency-shifts suggest that
the degree of anharmonicity remains approximately constant through a 
surface phonon band\cite{benedek92}. 

In conclusion, we have developed a simple model for the study of surface thermal expansion.
By applying it to Ag(111), we have demonstrated that the anomalously large surface thermal expansion
of Ag(111) can be attributed to a rapid softening of in-plane vibrational modes (related
to a rapid flattening of the corrugation of the interlayer potential) upon increasing
interlayer distances. 
We have shown that a similar scenario does not, however, lead to significant 
enhancement on Al(111).


\newpage
\Large
\begin{center}
{\bf Table 1}
\end{center}
\normalsize
\begin{center}
\begin{tabular}{|c|c|c|} \hline
 Quantity & Ag &  Al\\ \hline
 $a_0$(\AA)         &     4.06 {\it(4.09)}     & 3.94 {\it(4.05)}   \\ \hline
 $B$(MBar)          &     1.22 {\it(1.01)}     & 0.80 {\it(0.75)}   \\ \hline    
$B^\prime$          &     5.94 {\it(5.92)}     & 4.22 {\it(5.36)}    \\ \hline
$\gamma(T)$         &     2.39 {\it(2.46)}     & 2.12 {\it(2.18)}   \\ \hline
$\gamma(L)$         &     2.95 {\it(2.46)}     & 2.05 {\it(2.18)}   \\[0.2cm] \hline
\end{tabular}
\end{center}

Table 1: Calculated values of bulk properties of Ag and Al: $a_0$ = the equilibrium
lattice constant at $T=0$, $B$=bulk modulus, $B^\prime$ = pressure derivative of $B$,
$\gamma(T)$ = Gr\"uneisen parameter for transverse bulk mode along [111], with wave-vector 2/3
of the way between the zone-center and zone-edge, $\gamma(L)$ = Gr\"uneisen parameter for
longitudinal bulk mode at the same wave-vector. Experimental values at room temperature,
obtained from Reference \cite{moruzzi88}, are given in parentheses.

\newpage
\Large
\begin{center}
{\bf Figures}
\end{center}
\normalsize

\par\noindent
\psfig{figure=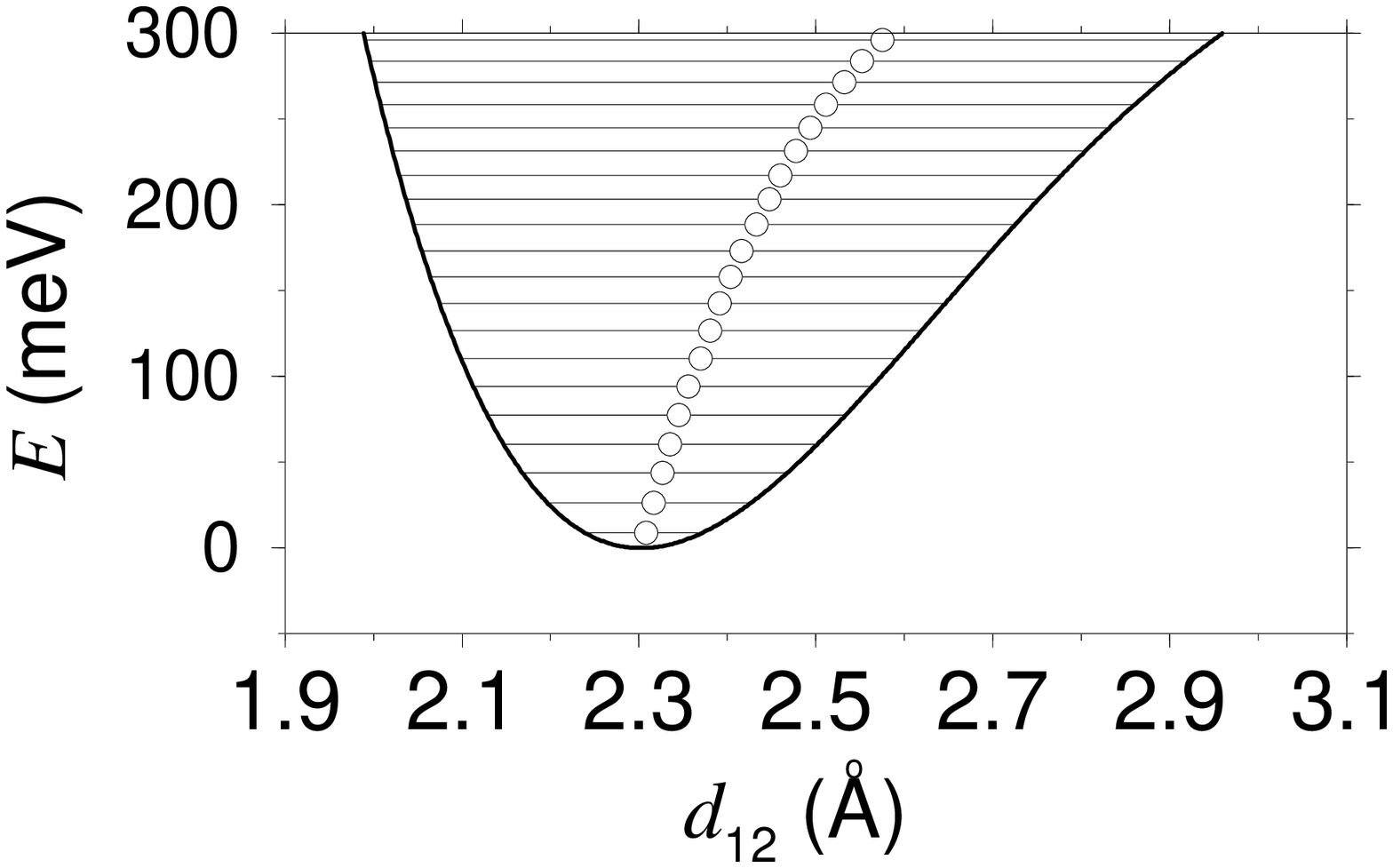}

Fig.1: Static interlayer potential between the two outermost layers
of Ag(111). The horizontal lines and open circles indicate the
energy eigenvalues $E_n$ and the mean displacements
$\langle d_{12}\rangle_n$ respectively of an Ag atom vibrating in this potential. 
\par\noindent

\newpage
\par\noindent
\psfig{figure=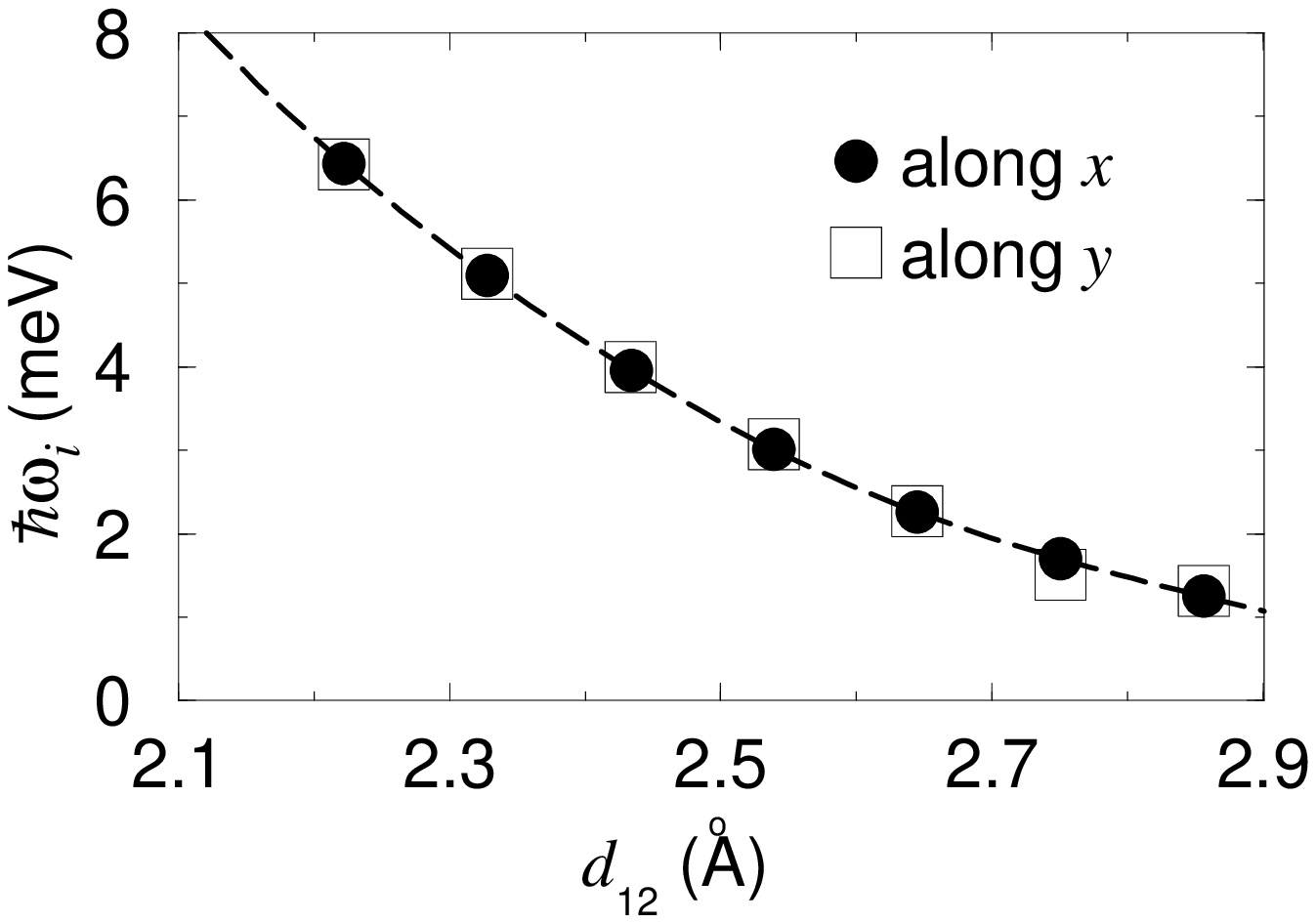}

Fig.2: Energy $\hbar\omega_i$ of the top-layer in-plane modes of
Ag(111), as a function of the first interlayer separation $d_{12}$. Circles
and squares indicate modes polarized along the $x$- and $y$- directions respectively.
\par\noindent
\
\par\noindent

\newpage

\psfig{figure=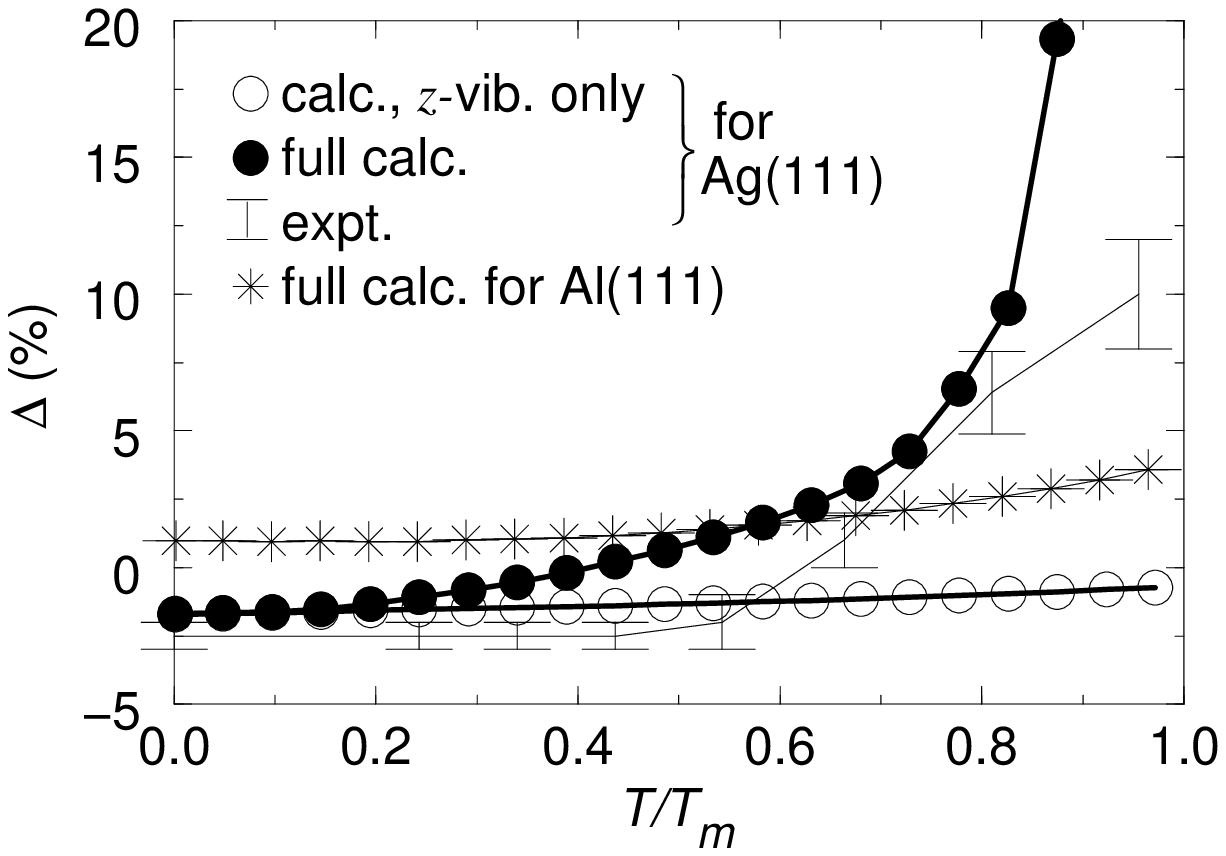}

Fig.3: Contraction/expansion of $d_{12}$ relative to $d_B$, as a function of
normalized temperature $T/T_m$ --
our calculations for Ag(111) with out-of-plane mode only (open circles),
and all three modes (filled circles); experiment$^1$ on
Ag(111) (solid line with error bars); our calculation for Al(111)
with all three modes (stars). $T_m$ is the bulk melting temperature 
($T_m^{Ag}$ =
1234 K, $T_m^{Al}$ = 933 K).
\par\noindent
\newpage

\par\noindent
\psfig{figure=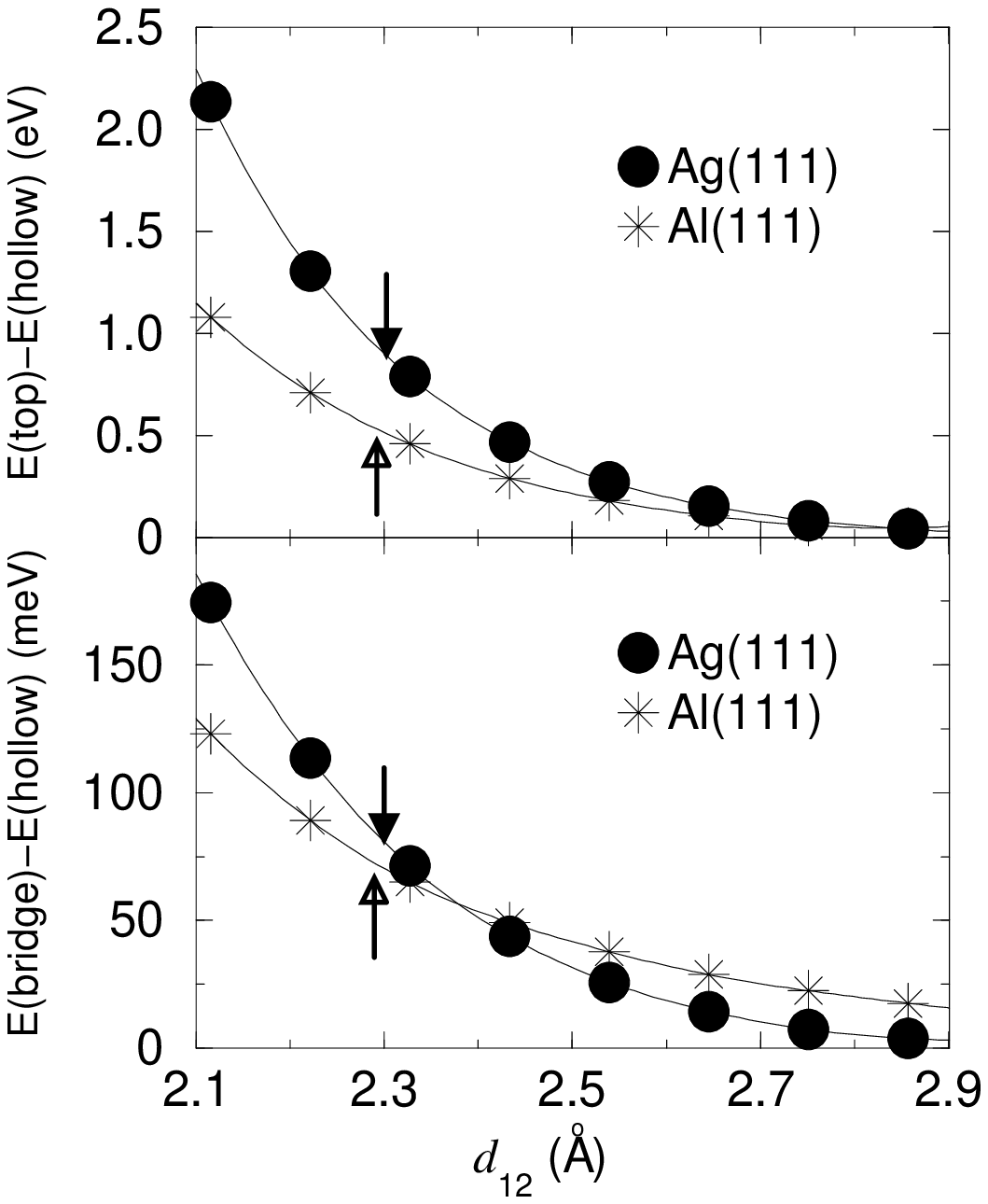}

Fig.4: Decrease in the corrugation of the interlayer potential with
increasing interlayer separation $d_{12}$: the  graphs show
the increase in surface energy when the outermost layer of atoms 
occupies an atop site or bridge site instead of the favored fcc hollow site,
for Ag(111) (filled circles) and Al(111) (stars). Arrows indicate the equilibrium
value (neglecting zero-point vibrations) of $d_{12}$ at $T$ = 0 K.


\begin{references}

\bibitem{statiris94}
P.~Statiris, H.C.~Lu and T.~Gustafsson, Phys.~Rev.~Lett. {\bf 72}, 3574 (1994).

\bibitem{explain}
Throughout this paper, when we refer to ``surface thermal expansion", it is implied
that we are referring to changes in interlayer separations. It is assumed that intralayer
separations are the same in the bulk and at the surface.

\bibitem{allen69}
R.E.~Allen and F.W.~de~Wette, Phys.~Rev. {\bf 179}, 873  (1969).

\bibitem{wilson71}
J.M.~Wilson and T.J.~Bastow, Surf.~Sci. {\bf 26}, 461  (1971 ).

\bibitem{dob73ma78}
L.~Dobrzynski and A.A.~Maradudin, Phys.~Rev. B {\bf 7}, 1207  (1973);
S.K.S.~Ma, F.W.~de~Wette and G.P.~Alldredge, Surf.~Sci. {\bf 78}, 598  (1978).

\bibitem{lewis95}
L.J.~Lewis, Phys.~Rev.~B {\bf 50}, 17 693   (1994).

\bibitem{frenken87}
J.W.M.~Frenken, F.~Huusen and J.F.~van der Veen, Phys.~Rev.~Lett. {\bf 58}, 401  (1987).

\bibitem{cao90}
Y.~Cao and E.~Conrad, Phys.~Rev.~Lett. {\bf 65}, 2808  (1990).


\bibitem{helgesen93}
G.~Helgesen, D.~Gibbs, A.P.~Baddorf, D.M.~Zehner and S.G.J.~Mochrie, Phys.~Rev.~B. {\bf 48},
15 320 (1993).

\bibitem{yang91}
L.~Yang and T. Rahman, Phys.~Rev.~Lett. {\bf 67}, 2327 (1991).

\bibitem{beaudet93}
Y.~Beaudet, L.J.~Lewis and M.~Persson, Phys.~Rev.~B {\bf 47}, 4127  (1993).


\bibitem{gonze91}
X.~Gonze, R.~Stumpf and M.~Scheffler,  Phys.~Rev.~B {\bf 43}, 8503 (1995).

\bibitem{ceperley80}
D.M.~Ceperley and B.J.~Alder, Phys. Rev. Lett. {\bf 45}, 566 (1980).


\bibitem{d23}
Over the temperature range we have considered, the relaxation of $d_{12}(T)$ 
changes by $< 0.3\%$, and $\alpha_S$ by $< 5\%$, upon allowing for the 
relaxation of $d_{23}$.

\bibitem{gschneider64}
K.A.~Gschneider,~Jr., in {\it Solid State Physics}, ed. by F.~Seitz and D.~Turnbull
(Academic, New York, 1964), Vol. 16, p. 27.

\bibitem{jayanthi}
C.S.~Jayanthi, E.~Tosatti and L.~Pietronero, Phys.~Rev.~B {\bf 31}, 
3456 (1985).


\bibitem{benedek92}
G.~Benedek and J.P.~Toennies, Phys.~Rev.~B {\bf 46}, 13 643  (1992).



\bibitem{au111expt}
J.V.~Barth, H.~Brune, G.~Ertl, and R.J.~Behm, Phys.~Rev.~B {\bf 42}, 9307 (1990);
K.G.~Huang, D.~Gibbs, D.M.~Zehner, A.R.~Sandy and S.G.J.~Mochrie, 
Phys. Rev. Lett. {\bf 65}, 3313 (1990).

\bibitem{pt111expt}
A.R.~Sandy, S.G.J.~Mochrie, D.M.~Zehner, G. Gr\"ubel, K.G.~Huang and D.~Gibbs,
Phys. Rev. Lett. {\bf 68}, 2192 (1992).

\bibitem{unpub}
S.~Narasimhan and M.~Scheffler, unpublished.

\bibitem{gustpriv}
T.~Gustafsson, private communication.

\bibitem{jiang91}
Q.T.~Jiang, P.~Fenter and T.~Gustafsson, Phys.~Rev.~B {\bf 44}, 5773  (1991).

\bibitem{moruzzi88}
V.L.~Moruzzi, J.F.~Janak and K.~Schwarz, Phys.~Rev. B {\bf 37}, 790  (1988).


\end{references}
\end{document}